\documentclass[prd,twocolumn,superscriptaddress,altaffilletter,amssymb,showpacs,nofootinbib]{revtex4}
\usepackage[dvips]{graphicx}
\usepackage{amsmath}
\newcommand{\be}{\begin{equation}}
\newcommand{\ee}{\end{equation}}
\newcommand{\bea}{\begin{eqnarray}}
\newcommand{\eea}{\end{eqnarray}}

\begin{document}


\title{Dynamics of Scalar Field Dark Matter With a Cosh-like Potential}

\author{Tonatiuh Matos} \email{tmatos@fis.cinvestav.mx}
\affiliation{Departamento de F{\'\i}sica, Centro de Investigaci\'on y
  de Estudios Avanzados del IPN, A.P. 14-740, 07000 M\'exico D.F.,
  M\'exico.}

\author{Jos\'e-Rub\'en
  Lu\'evano}\email{jrle@azc.uam.mx}\affiliation{$\dag$Departamento de
  Ciencias B\'asicas, UAM-A, C.P. 02200 M\'exico, D.F., M\'exico.}

\author{Israel
  Quiros}\email{israel@uclv.edu.cu}\affiliation{Universidad Central de
  Las Villas, Santa Clara, CP 54830, Cuba}

\author{L. Arturo
  Ure\~na-L\'opez$^*$}\email{lurena@fisica.ugto.mx}\affiliation{Departamento
  de F\'isica, DCI, Campus Le\'on, Universidad de Guanajuato, CP 37150,
  Le\'on, Guanajuato, M\'exico.}

\author{Jos\'e Alberto
  V\'azquez$^*$}\email{jvazquez@fis.cinvestav.mx}\affiliation{Departamento
  de F{\'\i}sica, Centro de Investigaci\'on y de Estudios Avanzados
  del IPN, A.P. 14-740, 07000 M\'exico D.F., M\'exico.}

\date{\today}

\begin{abstract} 
The dynamics of a cosmological model fueled by scalar field dark
matter with a cosh-like potential plus a cosmological constant is
investigated in detail. It is revealed that the late-time attractor is
always the de Sitter solution, and that, depending on the values of
the free parameters, the oscillating solution of the scalar field --
modeling cold dark matter -- mediates between some early stage (say, the
radiation-dominated solution) and the accelerating de Sitter attractor.
\end{abstract}

 
\maketitle

\section{Introduction}
One of the greatest mysteries of modern Cosmology doubtless is the
nature of dark matter (DM) (for a review see\cite{DMreviw}). We know
that more than 23\% of the matter of the universe is attractive, but
of unknown nature. The most accepted candidate are particles from the
Minimal Supersymmetric Standard Model (MSSM)\cite{MSSMreview}. This
paradigm has been proved at cosmological level with a great
success. For example, it predicts very well the formation and
clustering of galaxies\cite{StForreview}, and the micro-Kelvin
temperature fluctuations of the cosmic microwave background (CMB) of
the universe\cite{CMBreview}. These two predictions represent the main
achievements of the model. 

Nevertheless, in the last decade, some inconsistencies of the model
when calcullations are compared with the observations at galactic
level, have arosen. For example, amongst other problems, the predicted
number of satellite galaxies around big galaxies is much bigger than
the observed one\cite{satelites}, and the DM density profile at the
center of galaxies seems to be steeper than observed\cite{cusp}. 

Even though there have been several attempts to deal with these
inconsistencies\cite{inconsist}, one compelling possibility is the
hypothesis that the DM is a scalar field, more than a Standard Model
(SM) particle\cite{l9}. The idea is rooted on the fact that almost all
the unified theories of physics contain scalar fields as the simplest
geometrical objects. These scalar fields are often called Higgs,
Inflatons, Dilatons, Scalerons, Radions, etc. All of them are needed
to achieve mathematical and physical consistency of these
theories. With the discovery of the dark energy (DE), attempts to
relate scalar fields with the DE also have had some success
(see\cite{Copeland:2006wr} for a comprehensive review).

In Ref.~\cite{l9} some of us proposed that a scalar field rolling down
a convex self-interaction potential can be a reasonable candidate for
DM, calling this hypothesis as Scalar Field Dark Matter (SFDM). This
hypothesis has some nice features. For example, the SFDM does not need
extra hypothesis to explain the flat DM profile in the center of
galaxies\cite{argelia}, or the number of satellite galaxies around the
Milky Way\cite{further}. 

The SFDM hypothesis has been investigated for a number of
self-interaction potentials like the cosh-like \cite{wang,mul,tona},
and the quadratic ones \cite{tona,quad}, with the consequent discovery
of several exact solutions of cosmological interest. However, within
the cosmological context, a full and detailed study of the dynamics of
SFDM models, to uncover their relevant asymptotic properties, is still
desirable.

The main goal of the present paper is, precisely, to study within the
cosmological context and by means of the dynamical systems tools, the
asymptotic properties of the SFDM model driven by a cosh-like
potential. Several relevant cosmological solutions will be correlated
with such important dynamical systems concepts like past and future
attractors, signaling the way the cosmic dynamics transits from
early-time to intermediate and then to late-time asymptotic states. 

It will be demonstrated, in particular, that the SFDM model driven by
a cosh-like potential with a cosmological constant term added, is a
good candidate to correctly describe a cosmic dynamics whose fate is,
eventually, the inflationary de Sitter regime that can be correlated
with present accelerated stage of the expansion of the Universe. Due
attention will be paid to the late-time oscillatory solution that can
e associated with cold dark matter (CDM) behaviour in the model, and
that, inevitably, leads to the inflationary de Sitter attractor.

The paper has been organized as it follows. The relevant physical
features of the model are discussed in the next section, and then, in
section III, its mathematical features are specified. The details of
the (linear) dynamical systems study of the SFDM model with a
cosh-like potential, are given in section IV. The main results of the
study are summarized in section V. Finally, in section VII, the
physical discussion of the results and brief conclusions of the study
are provided. We use the natural system of units where $\kappa^2=8\pi
G=c=1$.

\section{Scalar Field Model with a cosh-like Potential}
In a cosmological context, scalar fields with a cosh-like
self-interaction potential have been studied, for instance,
in\cite{wang}. In that reference the following potential was
investigated:
\be 
V(\phi)=V_0 [\cosh(\lambda\phi)-1]^\beta \, , \label{sahni}
\ee 
where $\lambda$ and $\beta$ are free parameters. It was shown,
see also\cite{Turner:1983he}, that during the oscillatory phase the
virial theorem gives the following expression for the mean equation of
state:
\be 
\left\langle \omega_\phi \right\rangle = \left\langle
  \frac{\dot\phi^2-2V}{\dot\phi^2+2V} \right\rangle =
\frac{\beta-1}{\beta+1} \, . \label{mean}
\ee 
In consequence, for $\beta=1$ the scalar field behaves like
pressureless dust $\left\langle \omega_\phi \right\rangle = 0$. A
scalar field potential with this value of $\beta$ could therefore play
the role of cold dark matter in the universe. For $\beta<1/2$ this
potential is a good candidate for the quintessence
instead\cite{wang}.

In this paper we will focus our attention, precisely, in a cosh-like
potential(\ref{sahni}) with $\beta=1$ with a cosmological constant
term added to it. The latter ingredient is neccessary to include dark
energy in the model, as long as oscillations of the scalar field
around the minimum of $\cosh$-potential play the role of the CDM (in
fact SFCDM).

A reasonable physical assumption for the initial conditions is that
the energy density in the scalar field was comparable to that of
radiation at very early times, say, at the end of inflation
\cite{wang}. At such early times, the large value of $\phi$
makes the cosh-like potential(\ref{sahni}) behave as an exponential
one of the form $V(\phi) \propto \exp(-\beta \lambda \phi)$, and the
scalar field energy density decreases rapidly so that it remains
subdominant until recently. 

In later times the form of $V(\phi)$ changes to a power law, resulting
in rapid oscillations of the scalar ield around $\phi=0$. At this
stage the scalar field equation of state mimics that of CDM in
agreement with the present cosmological standard model paradigm. Due
to the addition of the cosmological constant term, the resulting
picture does indeed represent a unification of the dark matter and the
dark energy. The present model might be called $\Lambda$-SFCDM. 

The goal of this paper will be to describe the dynamical picture
produced by the cosh-like potential (\ref{sahni}) with $\beta=1$ and
with the addition of a cosmological constant term, to put the above
physical discussion on solid dynamical systems grounds.

\section{Mathematical Features of the Model}
In this paper we focus in a Friedmann-Robertson-Walker space-time with
flat spatial sections, filled with a mixture of two fluids: i) a
perfect fluid of ordinary matter with density $\rho_\gamma$ and
barotropic index $0 \leq \gamma \leq 2$ ($\gamma=1$ for dust,
$\gamma=4/3$ for radiation, etc.), and ii) a united $\Lambda$-SFCDM
component with energy density $\rho_\phi = \dot{\phi}^2/2 + V(\phi)$
and parametric pressure $p_\phi = \dot{\phi}^2/2 - V(\phi)$. The
relevant cosmological equations of the model are the following
\begin{subequations}
\bea 
\dot {H} &=& -\frac{1}{2} \left( \dot{\phi}^2 + \gamma \rho_\gamma
\right) \, , \\
\ddot{\phi} &=& -3H \dot{\phi} - \frac{dV}{d\phi} \, , \\
\dot{\rho}_\gamma &=& -3H \gamma \rho_\gamma \, , \label{feqs}
\eea
\end{subequations}
plus the Friedmann constraint:
\be 
H^2 = \frac{1}{3} \left( \rho_\gamma + \frac{1}{2} \dot{\phi}^2 + V
\right) \, . \label{friedmann}
\ee

As anticipated, here we deal with a cosh-like potential of the form
\be 
V(\phi) = V_0[ \cosh(\lambda \phi) - 1 ] + \Lambda \,
, \label{pot}
\ee 
where $\Lambda\geq 0$ is a positive cosmological constant, and the
parameter $\lambda$ is also a possitive quantity ($\lambda>0$). As it
has been discussed in the former section, the potential $V(\phi)$ in
(\ref{pot}) comprises both CDM and dark energy in a united
ansatz. Actually, for a vanishing $\Lambda=0$, this potential is a
particular case of (\ref{sahni}), when the free parameter $\beta=1$,
meaning that, once the oscillatory phase is attained, the scalar field
driven by the potential (\ref{pot}) behaves like dust, i. e., just
like CDM (see equation (\ref{mean})). Since in this model $\Lambda
\neq 0$, the scalar field energy density performs damped oscillations
around $V_{min} = \Lambda$, meaning that the CDM energy density --
accounted for by the oscillatory component -- decreases until,
eventually, the cosmological constant component dominates.

The potential $V(\phi)$ in (\ref{pot}) is a minimum at $\phi=0 , \;
\Rightarrow \; V_{min} = V(0) = \Lambda$. Additionally, in a natural
scenario for the cosmic dynamics driven by $V(\phi)$, the scalar field
$\phi$ runs from arbitrarily large negative values ($|\phi|\gg 1$), to
vanishing ones ($|\phi|\ll 1$). In consequence, at early times the
dynamics is driven by an exponential potential
\be 
|\phi| \gg 1 / \lambda \;\; \Rightarrow \;\; V(\phi) \approx \bar{V}_0
\; e^{-\lambda\phi}, \;\; \bar{V}_0 \equiv \frac{V_0}{2} \,
, \label{earlytime}
\ee 
whereas at late times it is associated with a power-law potential plus
a cosmological constant:
\be 
|\phi|\ll 1/ \lambda \;\; \Rightarrow \;\; V(\phi) \approx
\frac{m^2}{2} \phi^{2} + \Lambda, \;\; m^2 \equiv V_0 \lambda^{2} \,
. \label{latetime}
\ee 
The latter is just the quadratic potential that has been formerly
studied, for instance, in references \cite{tona,quad}.

\section{Dynamical Systems Study}
The dynamical systems tools offer a very useful approach to the study
of the asymptotic properties of the cosmological models
\cite{coley,Copeland:2006wr}. In order to be able to apply these tools
one has to (unavoidably) follow the steps enumerated below.
\begin{itemize}
\item First: to identify the phase space variables that allow writing
  the system of cosmological equations in the form of an autonomus
  system of ordinary differential equations (ODE). There can be
  several different possible choices, however, not all of them allow
  for the minimum possible dimensionality of the phase space.

\item Next: with the help of the chosen phase space variables, to
  build an autonomous system of ODE out of the original system of
  cosmological equations.

\item Finally (some times a forgotten or under-appreciated step): to
  indentify the phase space spanned by the chosen variables, that is
  relevant to the cosmological model under study.
\end{itemize}
After this recipe one is ready to apply the standard tools of the
(linear) dynamical systems analysis. 

\subsection{Autonomous System of ODE}
Let us to introduce the following dimensionless phase space variables
in order to build an autonomous system out of the system of
cosmological equations~(\ref{feqs}) and~(\ref{friedmann})\cite{wands}
\be  
x \equiv \frac{\dot{\phi}}{\sqrt{6}H} , \; y \equiv
\frac{\sqrt{V}}{\sqrt{3}H} \, . \label{variables}
\ee 
After this choice of phase space variables we can write the following
autonomous system of ordinary differential equations
\begin{subequations}
\bea
x^\prime &=& -\sqrt\frac{3}{2} \frac{\partial_\phi V}{V} y^2 - 3x +
\frac{3}{2}x (2x^2 + \gamma \Omega_\gamma ) \, , \label{eqx} \\
y^\prime &=& \sqrt{\frac{3}{2}} \frac{\partial_\phi V}{V} xy +
\frac{3}{2}y (2x^2 + \gamma\Omega_\gamma) \, , \label{eqy}
\eea
\end{subequations}
where a prime denotes derivative with respect to the time variable
$\tau\equiv\ln a$ (properly speaking the number of e-foldings of
expansion), and the dimensionless density parameter $\Omega_\gamma
\equiv \rho_\gamma /3H^2$ is given through the following expression
\be 
\Omega_\gamma = 1- x^2 - y^2 \, , \label{constraint}
\ee 
which is just a rewriting of the Friedmann constraint (\ref{friedmann}).

As long as one considers just constant and exponential
self-interaction potentials ($\partial_\phi V=0$ and $\partial_\phi
V = \rm{const}$, respectively), Eqs.~(\ref{eqx}) and~(\ref{eqy}) form a
closed autonomous system of ODE. However, if one wants to go further
and to consider a wider class of self-interaction potentials beyond
the exponential one -- as it is the case in the present study --, the
system of ODE~(\ref{eqx}) and~(\ref{eqy}) is not a closed system of
equations any more, since, in general, $\partial_\phi V$ is a function
of the scalar field $\phi$. 

A way out of this difficulty can be based on the method developed in
\cite{chinos}. In order to be able to study arbitrary self-interaction
potentials one needs to consider one more variable $v$, that is
related with the derivative of the self-interaction potential through
the following expression
\be 
v \lambda \equiv -\partial_\phi V/V =-\partial_\phi \ln V \,
. \label{s}
\ee 
Hence, an extra equation
\be 
v^\prime = -\sqrt{6} \lambda x v^2 (\Gamma-1) \, , \label{sn'}
\ee  
has to be added to the above autonomous system of equations. The
quantity $\Gamma \equiv V\partial_\phi^2 V/ (\partial_\phi V)^2$ in
Eq.~(\ref{sn'}) is, in general, a function of $\phi$. The idea
behind the method in\cite{chinos} is that $\Gamma$ can be written as a
function of the variable $v$, and, perhaps, of several constant
parameters. Indeed, for a wide class of potentials the above
requirement -- $\Gamma = \Gamma(v)$ --, is fulfilled. 

Let us introduce a new function $f(v)=v^2 (\Gamma(v)-1)$ so that
Eq.(\ref{sn'}) can be written in the more compact form
\be 
v^\prime = -\sqrt{6} \lambda x f(v) \, . \label{snn'}
\ee 
Eqs.~(\ref{eqx}), (\ref{eqy}), (\ref{constraint}), and~(\ref{snn'})
form a three-dimensional -- closed -- autonomous system of ODE
\begin{subequations}
\bea 
x^\prime &=& \sqrt\frac{3}{2} \lambda y^2v -3x +\frac{3}{2}x (2x^2 +
\gamma \Omega_\gamma) \, , \\
y^\prime &=& -\sqrt{\frac{3}{2}} \lambda xyv + \frac{3}{2}y (2x^2 +
\gamma\Omega_\gamma) \, , \\
v^\prime &=& -\sqrt{6} \lambda x f(v) \, , \;\; \Omega_\gamma = 1 -x^2
-y^2 \, , \label{ode}
\eea
\end{subequations}
that can be safely studied with the help of the standard dynamical
systems tools\cite{coley}. Notice the obvious symmetry of the
ODE's~(\ref{ode}) under the simultaneous change of sign of $\lambda$
and $v$: $\lambda\rightarrow -\lambda$, $v\rightarrow -v$.

An important limitation of the approach explained above is related
with the fact that, for potentials that vanish at the minimum -- such
as, for instance $V = V_0 (\cosh(\lambda\phi) -1)^\beta$, or $V =
m^2\phi^2/2$ --, the variable $v = -\partial_\phi V/(\lambda V)$ is
undefined at this important point, usually important for the late-time
dynamics. However, thanks to the non-vanishing cosmological constant
term $\Lambda$ in Eq.(\ref{pot}), the latter is not the case in the
present study, so that the approach of\cite{chinos} can be safely
applied.

In terms of the phase space variables $x$, $y$, and $v$, the following
relevant magnitudes: i) the deceleration parameter $q \equiv -(1 +
\dot{H}/H^2)$, and ii) the effective equation of state parameter of
the scalar field $\gamma_\phi \equiv 2\dot{\phi}^2/(\dot{\phi}^2 +
2V)$, can be written, respectively, as
\begin{subequations}
\bea
q &=& -1 + \frac{3}{2}(2x^2 + \gamma \Omega_\gamma) \, , \\
\gamma_\phi &=& \frac{2x^2}{x^2+y^2} \, . \label{parameters}
\eea 
\end{subequations}

\begin{figure}[ht!]
\begin{center}
\includegraphics[width=5cm,height=4cm]{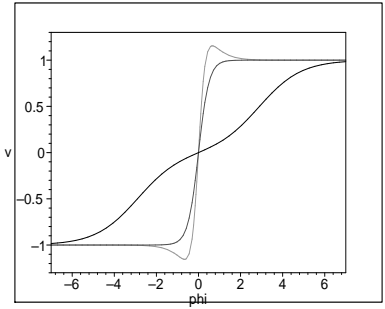}
\vspace{0.3cm}
\caption{A plot of function $v(\phi)$ for the chosen values of the
  parameters of the potential $\lambda=1$, $\alpha=10$ -- darker
  curve, $\alpha=1$ -- dark-to-gray curve, and $\alpha=0.5$ --
  soft-gray curve, respectively.}\label{fig1}
\end{center}
\end{figure}

\begin{figure}[t!]
\begin{center}
\includegraphics[width=4cm,height=3.5cm]{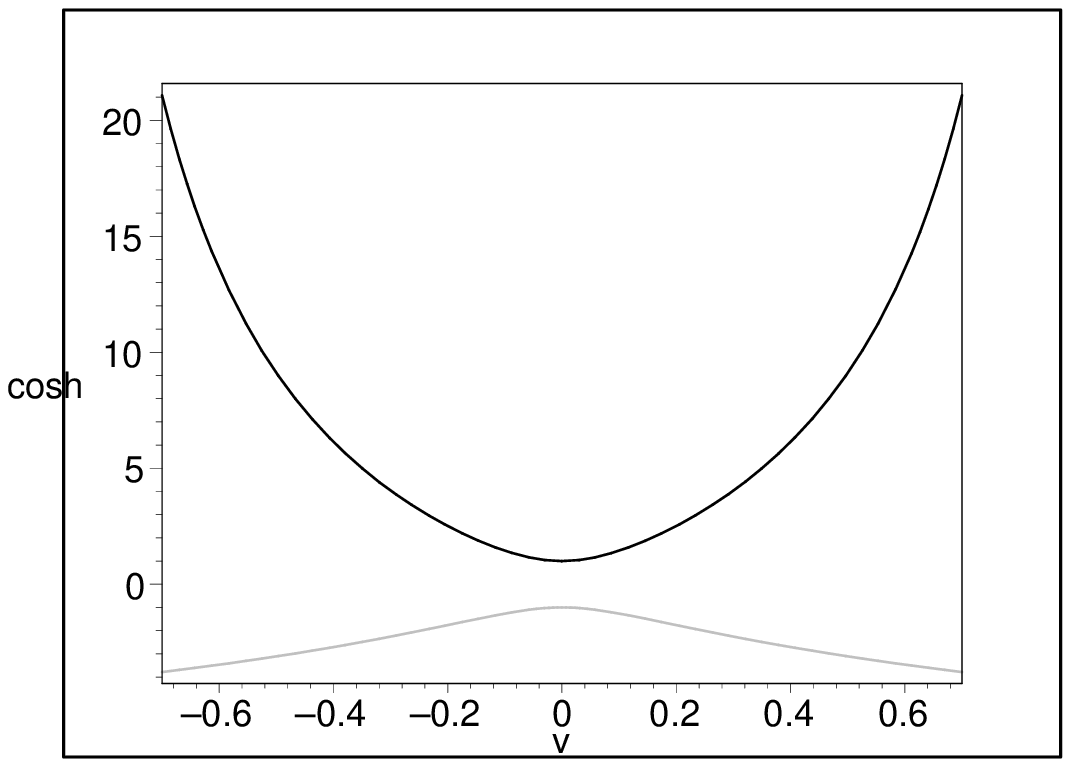}
\includegraphics[width=4cm,height=3.5cm]{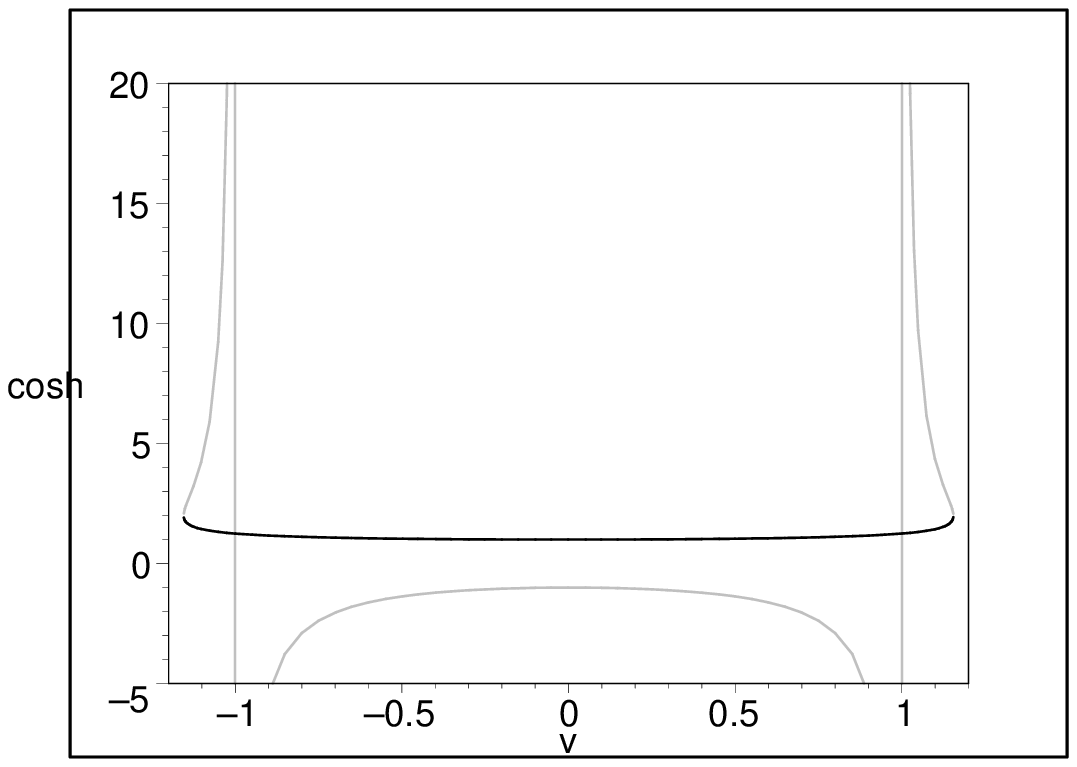}
\vspace{0.3cm}
\caption{A plot of function $\cosh(\lambda\phi)$ vs $v$, for the
  chosen values of the free parameters of the potential
  ($\lambda=1$,$\alpha=10$) -- left-hand panel, and
  ($\lambda=1$,$\alpha=0.5$) -- right-hand panel. Both branches of the
  cosh-function, the positive (dark curve) and the negative one (gray
  line) are shown. Notice that, for the second case ($\alpha<1$), the
  whole cosh-function behaviour is depicted by a union of part of the
  negative and the entire positive branches.}\label{fig1'}
\end{center}
\end{figure}

\begin{figure}[t!]
\begin{center}
\includegraphics[width=4cm,height=3.5cm]{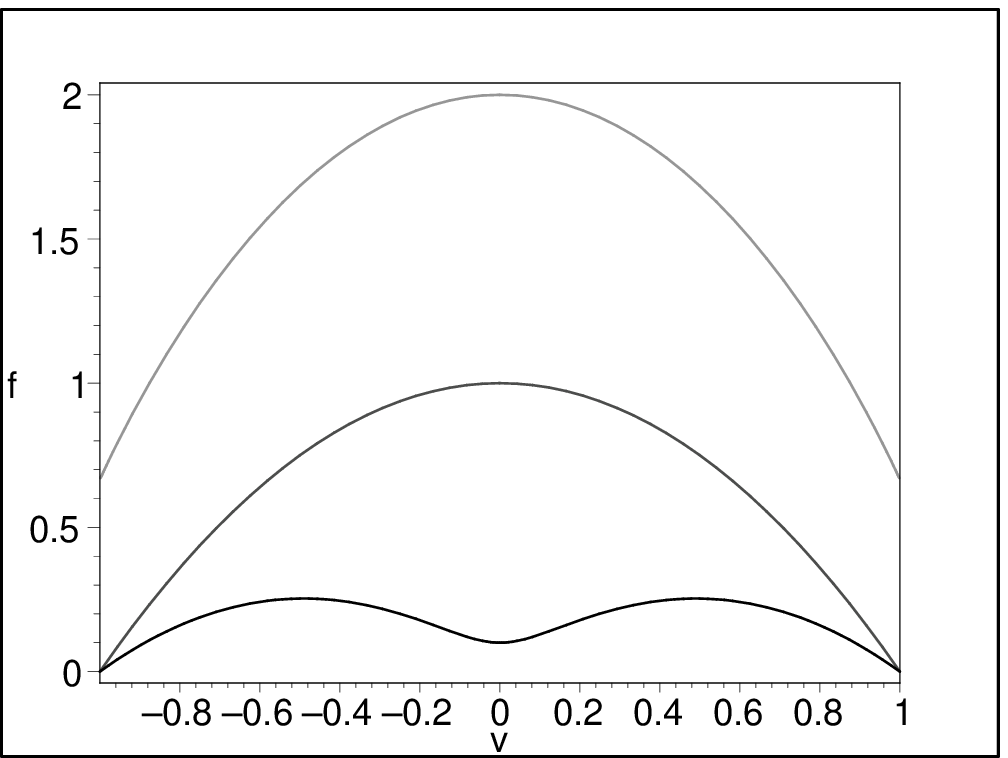}
\includegraphics[width=4cm,height=3.5cm]{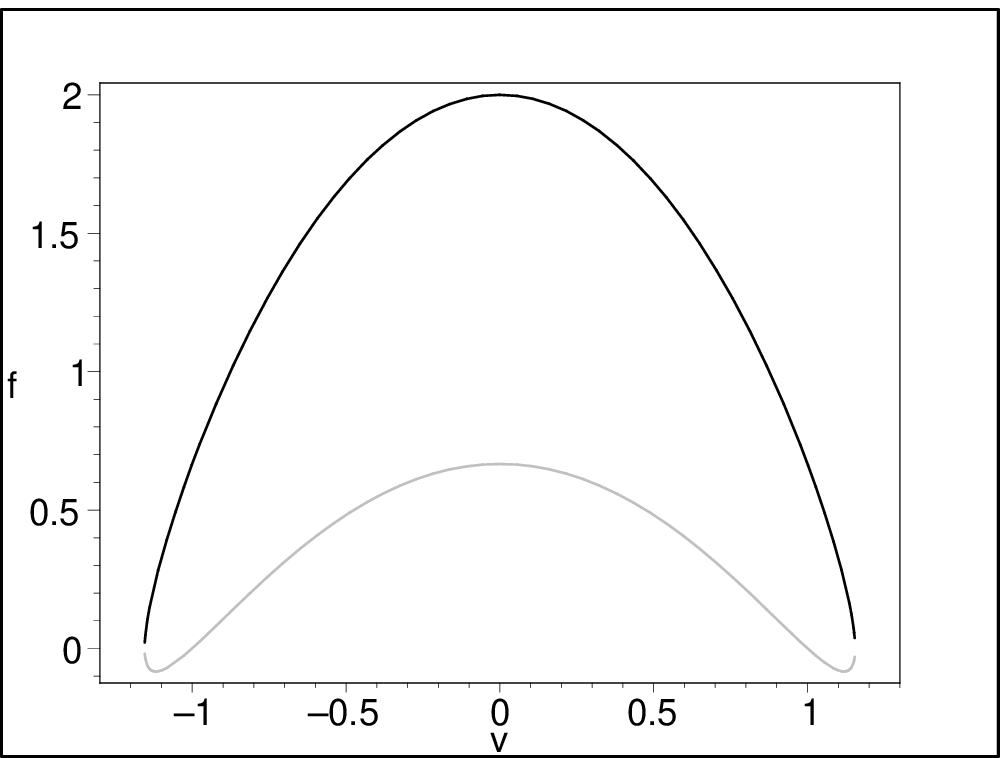}
\vspace{0.3cm}
\caption{A plot of function $f^+$ in Eq.(\ref{fv}) vs $v$, for the
  chosen values of the free parameters of the potential in
  Fig.\ref{fig1} -- left-hand panel. In the right-hand panel we show a
  plot of $f$ vs $v$ for $\alpha=0.5<1$ ($\lambda=1$). Both branches
  of $f(v)$ in Eq.(\ref{fv}), the positive (darker curve) and the
  negative one (gray line), are depicted.}\label{fig2}
\end{center}
\end{figure}

\subsection{Function $f(v)$ for the cosh-like Potential (\ref{pot})}
We now derive the function $f(v)$ in Eq.~(\ref{snn'}) (also present in
Eq.~(\ref{ode})), for the potential of interest here
(Eq.(\ref{pot})). According to the definition~(\ref{s}) for the
cosh-like potential, one has
\be
v(\phi) = \frac{\sinh(\lambda\phi)}{\cosh(\lambda\phi) -1 +\alpha} \,
, \label{vphi}
\ee
where $\alpha\equiv\Lambda/V_0$. From (\ref{vphi}) it follows, in
particular, that 
\be
\lim_{\phi \rightarrow \pm\infty} v(\phi) = \pm 1, \; \lim_{\phi
  \rightarrow 0} v(\phi) =0 \, .
\ee 
The value $\alpha=1$ is a critical one. Actually, the function~(\ref{vphi})
is an extremum whenever: 
\be
\cosh(\lambda\phi) = \frac{1}{1-\alpha}, \; \Rightarrow \; v^2_{ext} =
\frac{1}{\alpha(2-\alpha)} \, . 
\ee
The fact that the cosh function is always equal or bigger than unity
means that $v(\phi)$ is an extremum only if $0<\alpha< 1$. For $\alpha
\geq 1$ the function $v$ is, at least, a monotonically non-decreasing
function as the scalar field takes values in the interval $-\infty <
\phi < \infty$, therefore, in this case $v \in[-1,1]$. 

In Fig.~\ref{fig1} the function $v(\phi)$ is plotted for the following
sets of values of the parameters of the potential $V(\phi)$:
($\lambda=1, \alpha=10$) -- dark curve, ($\lambda=1, \alpha=1$) --
dark-to-gray curve, and ($\lambda=1, \alpha=0.5$) -- soft-gray curve,
respectively. As seen, in the latter case, since $\alpha<1$, $v(\phi)$
is a minimum at 
\be
\lambda\phi = -\text{arc}\cosh(\frac{1}{1 - \alpha}) =
-\text{arc}\cosh(2) \, ,
\ee
where $v_{min} = -1/\sqrt{\alpha(2 -\alpha)} = -\sqrt{4/3}$, while at
$\lambda\phi=\text{arc}\cosh(2)$ it is a maximum instead: $v_{max}
=\sqrt{4/3}$.

The function $v(\phi)$ in Eq.(\ref{vphi}) can be inverted to have the
cosh function written in terms of $v$,
\be 
\cosh(\lambda\phi)^\pm(v) = \frac{(\alpha -1) v^2 \pm\sqrt{1 +
    \alpha(\alpha - 2) v^2}}{1 - v^2},\label{coshv}
\ee
where the "$\pm$" signs refer to the two different branches of
$\cosh(\lambda\phi)(v)$. 

For values $\alpha \geq 1$, the mathematical requirement that the
cosh-function has to be, neccessarily, bigger than or equal to the
unity, leads to the following condition
\be 
\alpha v^2 \pm \sqrt{1+\alpha(\alpha-2)v^2} \geq 1, \; \Rightarrow \;
|v| \leq 1 \, ,
\ee 
so that, as already mentioned before, only values of $v$ obeying
$|v| \leq 1$ can be considered. This point is relevat when discussing
the definition of the phase space since $v$ is one of the variables
spanning the phase space of the model under consideration here. For
$0 < \alpha <1$ the reasoning is not so simple, however, it is clear
from the former analysis that, in this case, $|v| \leq |v_{ext}| =
1/\sqrt{\alpha(2 - \alpha)}$. 

In order to illustrate the discussion above, in Fig.~\ref{fig1'} we
plotted the cosh-function vs $v$ for two sets of values of the
parameters of the potential: i) ($\lambda=1,\alpha=10$) -- left-hand
panel, and ii) ($\lambda=1,\alpha=0.5$) -- right-hand panel,
respectively. In both cases the darker curve is for the positive
branch ("+" sign in Eq.~(\ref{coshv})), while the gray line is for the
negative branch ("-" sign in Eq.~(\ref{coshv})) instead. 

Notice from the left-hand panel of Fig.~\ref{fig1'} ($\alpha=10>1$),
that only the positive branch actually depicts the right (whole)
cosh-function behaviour. In the right-hand panel ($\alpha=0.5<1$),
instead, the whole cosh-function behaviour can be covered by a union
of that part of the negative branch to the left of the vertical
asymptote $v=-1$, starting at infinitely large values of
$\cosh(\lambda\phi)$ (i. e., infinitely large negative values of
$\phi$) until the gray curve meets the dark one (the positive branch)
at the union point at $v = v_{min} =-1/\sqrt{\alpha(2 - \alpha)}$. 

The same is true for positive values of $\phi$: the right
cosh-function behaviour is depicted by a union of that part of the
negative branch to the right of the vertical asymptote at $v = 1$,
starting at infinitely large positive values of $\phi$, until it
joints the positive branch at $v = v_{max} =1/\sqrt{\alpha(2
  -\alpha)}$. The range of intermediate-to-small values of $\phi$
(including the beginning at $\phi=0$, where the potential $V(\phi)$ in
Eq.~(\ref{pot}) is a minimum), is completely covered by the positive
branch of the cosh-function~(\ref{coshv}).

For the potential~(\ref{pot}) the function $f(v)$ appearing in
Eqs.~(\ref{snn'}) and~(\ref{ode}), can be written in the following
way
\be 
f^\pm(v) = \frac{\pm\left( 1 - v^2 \right) \sqrt{1 +\alpha(\alpha -2)
    v^2}}{\alpha -1 \pm\sqrt{1 + \alpha(\alpha -2) v^2}} \,
, \label{fv}
\ee 
where the $\pm$ signs refer to the positive and the negative branches
of the function $f(v)$ (directly related with the positive and the
negative branches of the cosh-function in Eq.~(\ref{coshv})),
respectively. In Eq.~(\ref{fv}) the choice of the "+" or the "-" sign
is to be made simultaneously in the numerator and in the denominator
on the right-hand-side. Hence, for instance, for the positive branch of
$f(v)$ one has 
\be
f^+(v) = \frac{+(1 - v^2) \sqrt{...}}{\alpha -1 +\sqrt{...}}, \;\;
\text{etc.}
\ee
It has to be emphasized that, while for values of the free parameter
$\alpha\geq 1$ the positive branch $f^+(v)$ in Eq.~(\ref{fv}) is enough
to depict the whole dynamics, for $0 < \alpha < 1$, instead, one needs
to contemplate both branches $f^\pm(v)$. 

Actually, in the later case the piece of the dynamics in the
$v$-interval 
\be 1 < |v| \leq \frac{1}{\sqrt{\alpha(2 - \alpha)}} \, ,
\ee 
is uncovered by the correponding values of $f^-(v)$ in Eq.~(\ref{fv}):
$f^-(|v|>1)$. The rest of the dynamics -- including the late-time
behaviour -- is uncovered by the piece of the positive branch liying
in the interval $|v|\leq 1$: $f^+(|v|\leq 1)$. In consequence, for $0
< \alpha < 1$, the whole dynamics is uncovered by
\be 
f_{0 < \alpha < 1} = f^-(|v|>1) \cup f^+(|v|\leq 1) \,
. \label{alphaleq1}
\ee

To summarize, for $\alpha \geq 1$ the cosmic dynamics driven by the
potential (\ref{pot}) can be associated with the following
3-dimensional (compact) phase space, spanned by the variables $x$,
$y$, and $v$ (we take into account only expanding cosmologies $H \geq 0$)
\bea
\Psi_{\alpha \geq 1} &=& \left\{ (x,y,v): 0 \leq x^2 + y^2 \leq 1,
\right. \nonumber \\
&&\; \left. |x| \leq 1, \; 0 \leq y \leq 1, \; |v| \leq
1 \right\}, \label{phasespace}
\eea 
and we have to worry only about the positive branch of $f(v)$ in
Eq.(\ref{fv}). Meanwhile, for $0 < \alpha < 1$, the 3-dimensional
(compact) phase space is given by 
\bea
\Psi_{0 < \alpha < 1} &=& \left\{ (x,y,v): 0 \leq x^2 + y^2 \leq 1,
\right. \nonumber\\
&& \; \left. |x| \leq 1, \; 0 \leq y \leq 1, \; |v| \leq
\frac{1}{\sqrt{\alpha (2 - \alpha)}} \right\}, \label{phasespace'}
\eea 
and, depending on the piece of the dynamics one is interested in, one
has to rely either on the negative branch of Eq.(\ref{fv})
$f^-(|v|>1)$ (early times-to-intermediate dynamics), or on the
positive brach instead $f^+(|v| \leq 1)$ (late-time dynamics), the
whole dynamics being described by $f_{0<\alpha<1}$ in
Eq.(\ref{alphaleq1}).

In what follows, depending on the value of parameter $\alpha =
\Lambda/V_0$, we shall look for fixed points of the autonomous system
of ODE~(\ref{ode}), with $f(v)$ given, either by: i) the positive
branch of equation (\ref{fv}), within the phase space
$\Psi_{\alpha \geq 1}$ defined in Eq.~(\ref{phasespace}), whenever
$\alpha \geq 1$, or ii) by $f_{0 < \alpha < 1}$ defined in
Eq.~(\ref{alphaleq1}), within the phase space $\Psi_{0 < \alpha < 1}$ defined in Eq.~(\ref{phasespace'}), if $0 < \alpha < 1$.

\begin{table*}[t!]\caption[crit]{Properties of the fixed points for
    the autonomous system~(\ref{ode}).}
\begin{tabular}{@{\hspace{4pt}}c@{\hspace{14pt}}c@{\hspace{14pt}}c@{\hspace{14pt}}
c@{\hspace{14pt}}c@{\hspace{14pt}}c@{\hspace{14pt}}c@{\hspace{14pt}}c@{\hspace{14pt}}
c@{\hspace{14pt}}c@{\hspace{14pt}}c@{\hspace{14pt}}c}
\hline \hline \\[-0.3cm]
$P_i$ & $x$ & $y$ & $v$ & Existence & $\Omega_\gamma$ &
$\Omega_{\phi}$ & $\gamma_\phi$ & $q$ \\ [0.1cm] \hline \\ [-0.2cm]
$P_1$ & $0$ &$0$ &$v$ & Always &1 &$0$ & undefined &$-1 +
\frac{3\gamma}{2}$ \\[0.2cm]
$P_2$ & $1$ & $0$ & $1$ & " & $0$ & $1$ & $2$ & $2$ \\[0.2cm]
$P_3$ & $\frac{\lambda}{\sqrt6}$ & $\sqrt{1-\frac{\lambda^2}{6}}$ &
$1$ & $\lambda \leq \sqrt{6}$ & $0$ & $1$ & $\frac{\lambda^2}{3}$ &
$-1 + \frac{\lambda^2}{2}$ \\[0.2cm]
$P_4$ & $\sqrt\frac{3}{2} \frac{\gamma}{\lambda}$ &
$\sqrt\frac{3\gamma(2 - \gamma)}{2\lambda^2}$ & $1$ & $\lambda \geq
\sqrt{3\gamma}$ & $1 - \frac{3\gamma}{\lambda^2}$ &
$\frac{3\gamma}{\lambda^2}$ & $\gamma$ & $-1 + \frac{3\gamma}{2}$
\\[0.2cm]
$P_5$ & $0$ & $1$ & $0$ & Always & $0$ & $1$ & $0$ & $-1$ \\[0.4cm]
\hline\hline
\end{tabular}\label{tab1}
\end{table*}

\begin{table*}\caption[eigenv]{Eigenvalues of the linearization
    matrices corresponding to the fixed points in table \ref{tab1}. We
    have used the following definition $R \equiv \sqrt{11 \gamma^2 -
      28\gamma +12 \left(1 + 2\gamma^2 \left[ 3\gamma -2 +12(1 -
          \gamma) \gamma^2/ \lambda^2 - \gamma^2 (1 - 3
          \gamma^2/\lambda^2 ) \right]/ \lambda^2 \right)}$.}
\begin{tabular}{@{\hspace{4pt}}c@{\hspace{14pt}}c@{\hspace{14pt}}c@{\hspace{14pt}}
c@{\hspace{14pt}}c@{\hspace{14pt}}c@{\hspace{14pt}}c}
\hline \hline \\[-0.3cm]
$P_i$ & $\lambda_1$ & $\lambda_2$ & $\lambda_3$
\\[0.1cm]\hline\\[-0.2cm]
$P_1$ & $-3 +3\gamma/2$ & $3\gamma/2$ & $0$ \\[0.2cm]
$P_2$ & $6 -3\gamma$ & $3-\sqrt{3/2} \lambda$ & $\sqrt{6}\lambda$
\\[0.2cm]
$P_3$ & $-3 +\lambda^2/2$ & $\lambda^2 -3\gamma$ & $\lambda^2$
\\[0.2cm]
$P_4$ & $-3(2 -\gamma)/4 + \sqrt{3}R/4$ & $-3(2 - \gamma)/4
-\sqrt{3}R/4$ & $3\gamma$ \\[0.2cm]
$P_5$ & $-3\gamma$ & $-3(1 -\sqrt{1 -4\lambda^2/\alpha})/2$ & $-3(1
+\sqrt{1 -4\lambda^2/\alpha})/2$ \\[0.4cm]
\hline \hline
\end{tabular}\label{tab2}
\end{table*}

\subsection{Equilibrium Points and Stability}
The fixed points of the autonomous system of ODE~(\ref{ode}), in the
phase space $\Psi_{\alpha \geq 1}$ defined by Eq.~(\ref{phasespace})
or in $\Psi_{0 < \alpha < 1}$ defined by Eq.~(\ref{phasespace'}), are
listed in Table~\ref{tab1}, while the eigenvalues of the corresponding
linearization matrices are shown in Table~\ref{tab2}.

In the case when the parameter $0 < \alpha < 1$, by looking at the
definition of the function $f(v)$ in Eq.~(\ref{fv}), it might seem
that, in addition to the critical points in Table~\ref{tab1}, there
can be also equilibrium points associated with the values $v = \pm
1/\sqrt{\alpha(2 - \alpha)}$. However, if one looks at Fig.~\ref{fig2}
(right-hand panel), one can see that at $v = \pm 1/\sqrt{\alpha(2 -
  \alpha)}$ the derivative of the function $f_{0 < \alpha < 1}$ in
Eq.~(\ref{alphaleq1}), is undefined, so that the linear approach
undertaken in this investigation is not valid any more. This means
that the above points in phase space might not be actual critical
points, so that we shall not include them in our analysis.

Existence of the matter-dominated solution (equilibrium point $P_1$ in
Table~\ref{tab1}), is independent on the value of the variable $v$,
meaning that this phase of the cosmic evolution may arise at
early-to-intermediate times, as well as at late times. As seen from
Table~\ref{tab2}, since in this case the two non vanishing eigenvalues
of the linearization matrix are of oposite sign, the matter-dominated
solution is always a saddle equilibrium point of Eqs.~(\ref{ode}).

Equilibrium points $P_2$, $P_3$, and $P_4$, are associated with early
time dynamics since, according to Eq.~(\ref{vphi}), $v=1$ is correlated
with infinitely large values of the variable $\phi$. Besides, in this
case, \be
\partial_\phi \ln V = -\lambda , \; \Rightarrow \; V = \bar{V}_0
e^{-\lambda \phi}\, 
\ee
in correspondence with the limit of potential~(\ref{pot}) at large
$\phi$'s. This confirms that equilibrium points with $v=1$, indeed
correspond to early-times dynamics in the model with the cosh-like
potential~(\ref{pot}). The main porperties of these equilibrium points
($P_2-P_4$), together with those of the matter-dominated phase $P_1$,
have been discussed in\cite{wands}. These can be summarised as follows
(the properties of the matter-dominated solution $P_1$ have been
discussed above):
\begin{itemize}
\item The kinetic energy-dominated solution (point $P_2$ in
  Table~\ref{tab1}), is always decelerating. For $\lambda \leq
  \sqrt{6}$ it is the past attractor for any phase space trajectory.

\item The scalar field-dominated solution (equilibrium point $P_3$),
  exist whenever $\lambda \leq \sqrt{6}$, and it is inflacionary for
  $\lambda < \sqrt{2}$ (decelerating otherwise). It is always a saddle
  point in the phase space $\Psi$ defined in Eq.~(\ref{phasespace}).

\item The scaling solution (fixed point $P_4$), exists whenever
  $\lambda \geq \sqrt{3\gamma}$. This solution is correlated with
  decelerated expansion of the universe for $\gamma \geq 2/3$ (it is
  inflationary otherwise). The $\phi$-field mimics matter with a
  barotropic index $\gamma_\phi = \gamma$, while their energy
  densities scale as 
\be
\frac{\Omega_\phi}{\Omega_\gamma} = \frac{1}{\lambda^2/3\gamma -1} \,
.
\ee
It is always a saddle point in $\Psi$.

\item The late-time dynamics driven by potential~(\ref{pot}) is
  correlated with the minimum $V_{min} = \Lambda$ at $\phi=0$ -- and
  its neigbourhood --, which, in the phase space $\Psi$, is depicted
  by   the equilibrium point with $v=0$ in Table~\ref{tab1} (point
  $P_5$). This point corresponds to the inflationary de Sitter
  solution $3H^2 = \Lambda$. Since the real parts of the eigenvalues
  of the linearization matrix for $P_5$ are all negative (see
  Table~\ref{tab2}), the de Sitter solution is always the future
  attractor for any phase space trajectory.
\end{itemize}

For values of the parameter $\lambda > \sqrt{\alpha}/2$, the de Sitter
attractor is a spiral fixed point and, whenever $\alpha < 24$ both, the
past attractor (point $P_2$), and the future (late-time) attractor
(point $P_5$) co-exist, meaning that the corresponding orbits in the
phase space will be repelled by the kinetic energy-dominated solution
and will be eventually attracted to the de Sitter solution.

In this case ($\lambda > \sqrt{\alpha}/2$), the linear perturbations
of the scalar field perform damped oscillations around the minimum of
the potential at $\phi=0 \; \Rightarrow \; v = 0$, that are
characterized by cyclic frequency 
\be 
\omega = \frac{3}{2} \sqrt{\frac{4\lambda^2}{\alpha} -1} \, .
\ee
For $\lambda^2 > \alpha/4$, the general solution for the evolution of
linear perturbations $\delta x_i = (\delta x, \delta y,\delta v)$ in
the neighbourhood of the minimum of the potential (equilibrium point
$P_5$ in tables Tables~\ref{tab1} and~\ref{tab2}), can be written as
\bea
\delta x_i &=& a_{i1} e^{\lambda_1 \tau} +a_{i2} e^{\lambda_2 \tau} +
a_{i3} e^{\lambda_3 \tau} \nonumber \\
&=& a_{i1} e^{-3\gamma \tau} + e^{-3\tau/2} \left( a_{i2} \,
  e^{i\omega \tau} + a_{i3} \, e^{-i\omega \tau} \right) \,
, \label{perts}
\eea 
where $\lambda_1$, $\lambda_2$, and $\lambda_3$ are the eigenvalues of
the linearization matrix around $P_5$.

The latter oscillatory behaviour with frequency $\omega$ is what,
according to Refs.\cite{wang,further,mul}, can be associated with cold
dark matter because the amplitudes of the perturbations decrease at a
rate $\propto \exp{(-3\tau/2)}$.  The corresponding \emph{mean} energy
density $\left\langle \rho_\phi \right\rangle$ then dilutes at a rate
$\propto a^{-3}$ with an effective equation of state $\left\langle
  \omega_\phi \right\rangle =0$ (see Eq.~(\ref{mean})).

\begin{figure}[ht!]
\begin{center}
\includegraphics[width=4cm,height=3.5cm]{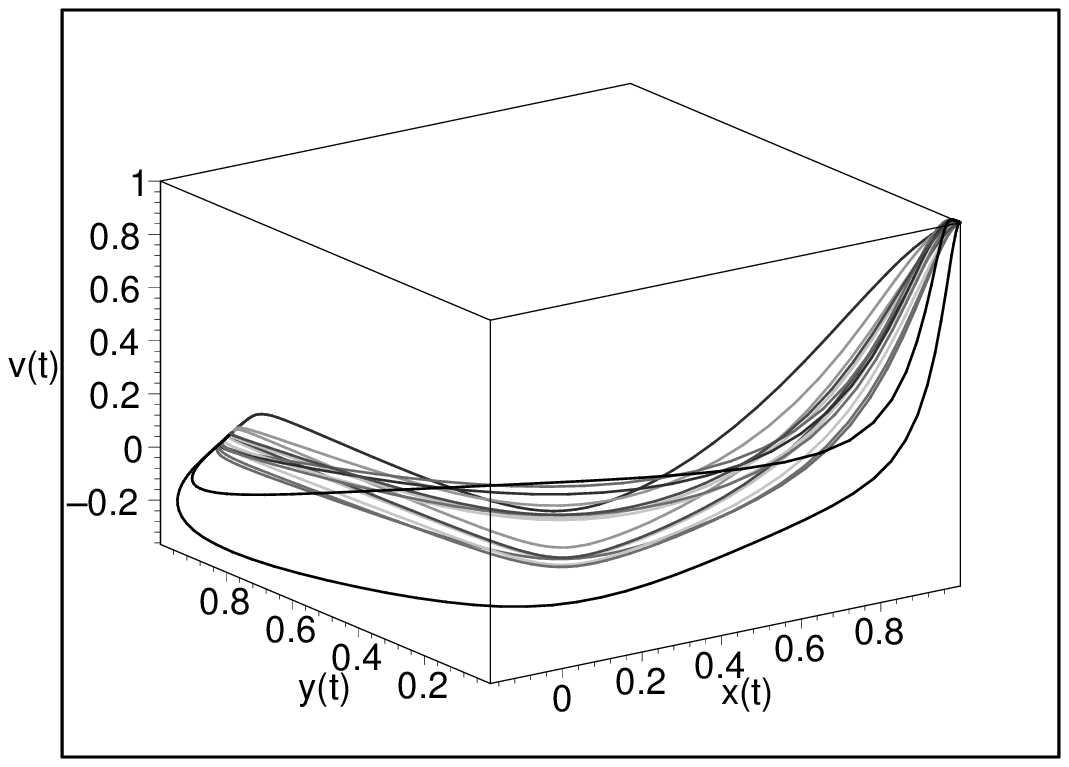}
\includegraphics[width=4cm,height=3.5cm]{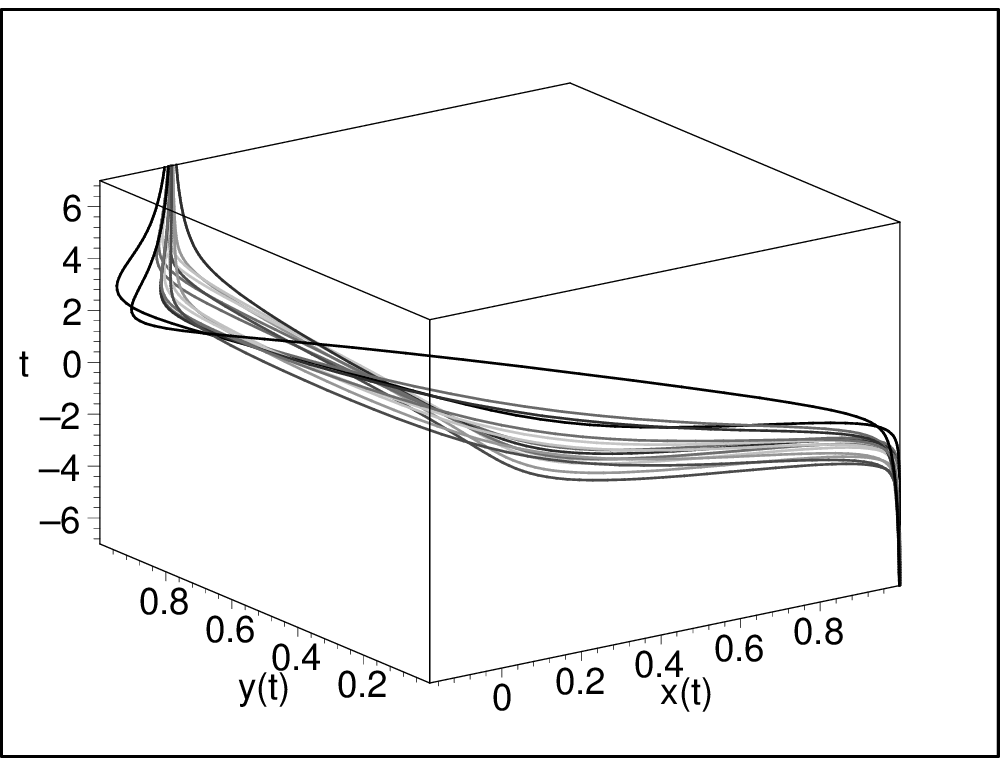}
\includegraphics[width=4cm,height=3.5cm]{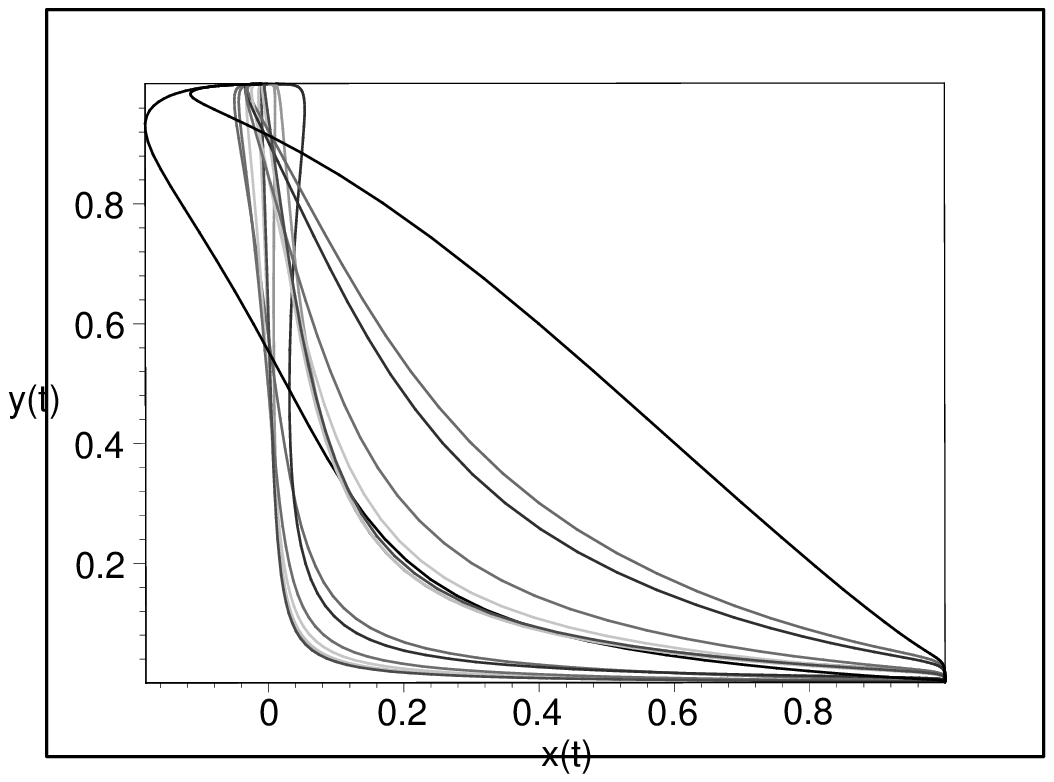}
\vspace{0.3cm}
\caption{Orbits of the autonomous system of ODE (\ref{ode}) with
  function $f(v)$ in Eq.~(\ref{fv}), and the corresponding flux in time
  $\tau$ -- upper panels --, for the chosen values of the parameters
  ($\lambda=2$, $\alpha=10$, $\gamma=4/3$-- background radiation), for
  different sets of initial conditions. In the lower panel we show the
  projection of the orbits into the phase plane $(x,y)$. Notice that
  the kinetic energy-dominated solution $(x,y,v) = (1,0,1)$ is the past
  attractor, while the de Sitter solution $(0,1,0)$ is the future
  attractor. The radiation-dominated solution $(0,0,v)$ is always a
  saddle fixed point for any given $v$.}\label{fig3}
\end{center}
\end{figure}

\begin{figure}[t!]
\begin{center}
\includegraphics[width=4cm,height=3.5cm]{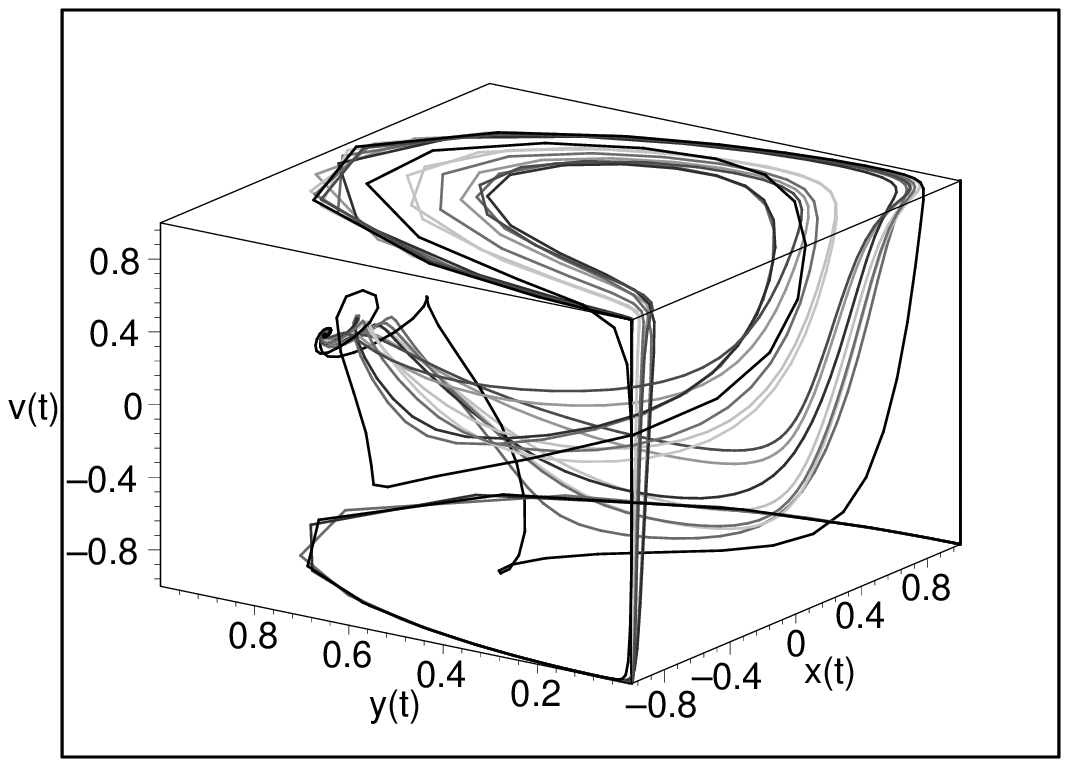}
\includegraphics[width=4cm,height=3.5cm]{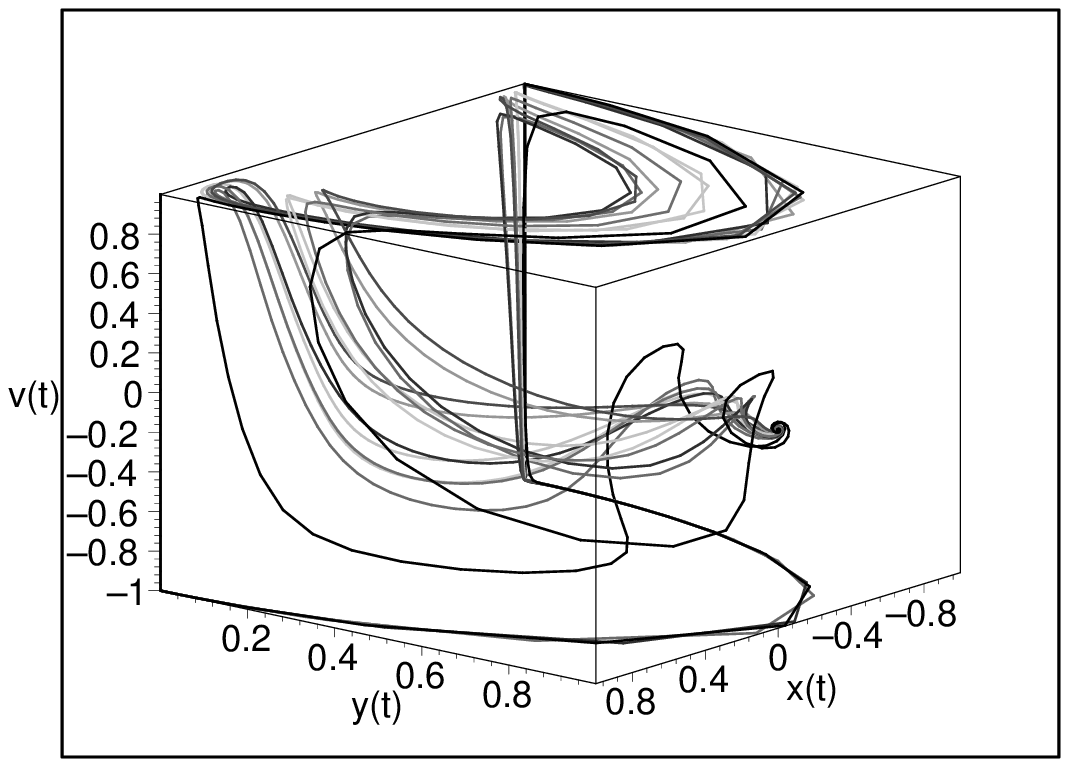}
\includegraphics[width=4cm,height=3.5cm]{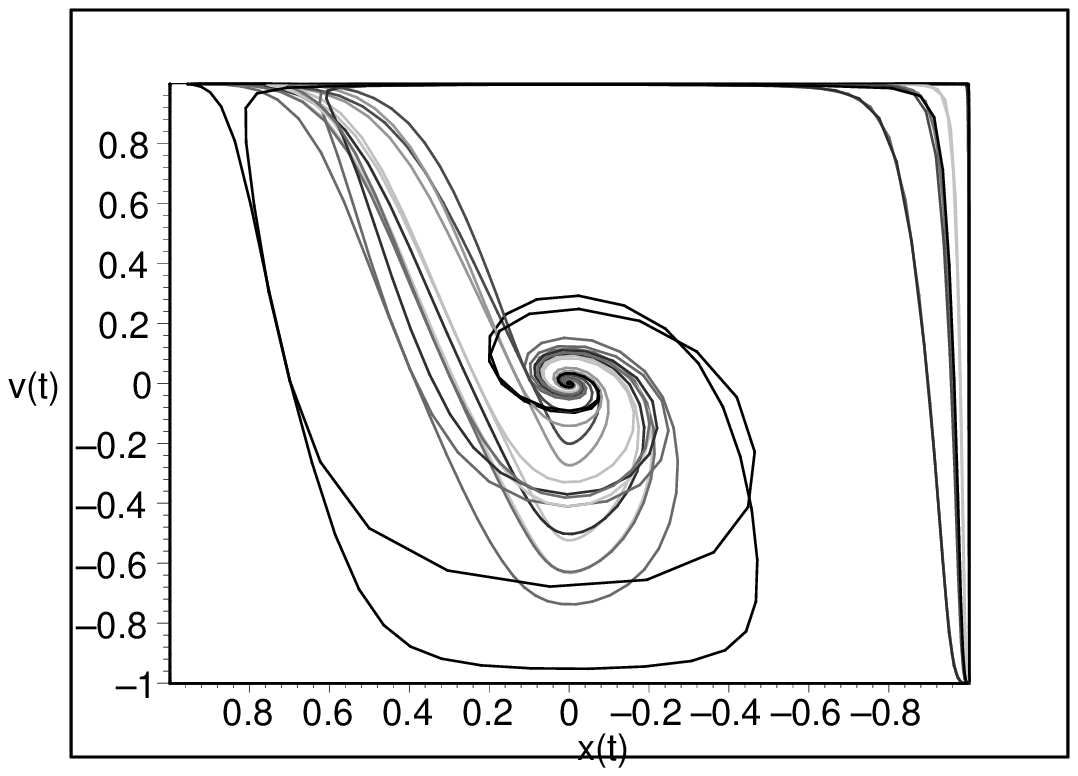}
\vspace{0.3cm}
\caption{Different perspectives of the local late-time behaviour of
  the orbits of the autonomous system of ODE (\ref{ode}), with
  function $f(v)$ in Eq.~(\ref{fv}), for different values of the
  parameters $\lambda=5$, and $\alpha=3$ ($\gamma=4/3$-- background
  radiation) -- upper panels. In the lower panel we show the
  projection of the orbits into the phase plane $(x,v)$. At late times
  the phase space trajectories coil around the de Sitter segment
  $\{(0,y,0) : 0 \leq y \leq 1\}$ until, eventually, they reach the
  inflationary de Sitter attractor $(0,1,0)$.}\label{fig4}
\end{center}
\end{figure}

The above results are illustrated in Fig.~\ref{fig3} and
Fig.~\ref{fig4} for a fixed value $\gamma = 4/3$. In particular, the
oscillating solution is depicted in the latter figure, where it is
apparent the way the orbits of the ODE~(\ref{ode}), at late times,
coil around the segment $\{(x,y,v) =(0,y,0) : 0 \leq y \leq 1\}$
until, eventually they reach the inflationary de Sitter attractor
$P_5=(0,1,0)$. The oscillating solution arises due to the choice of
the free parameters in Fig.~\ref{fig4}: $\lambda=5, \; \alpha =3$,
that obey the constraint $\lambda > \sqrt{\alpha}/2$. 

\begin{figure}[t!]
\begin{center}
\includegraphics[width=4cm,height=3.5cm]{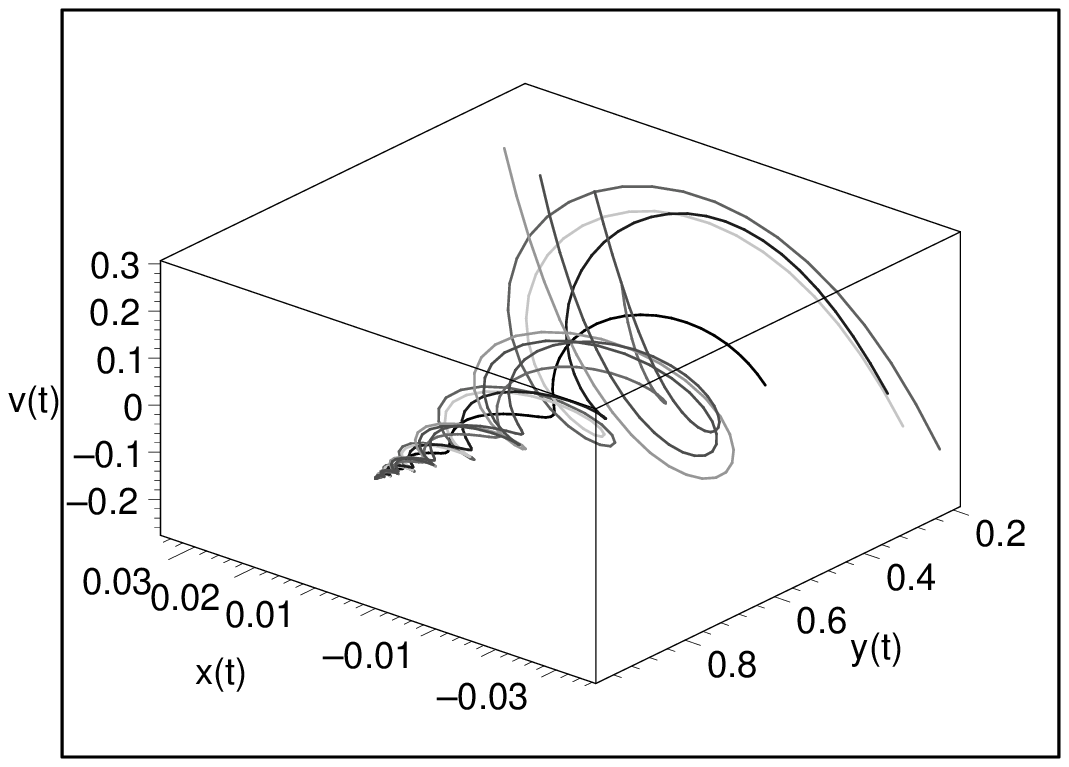}
\includegraphics[width=4cm,height=3.5cm]{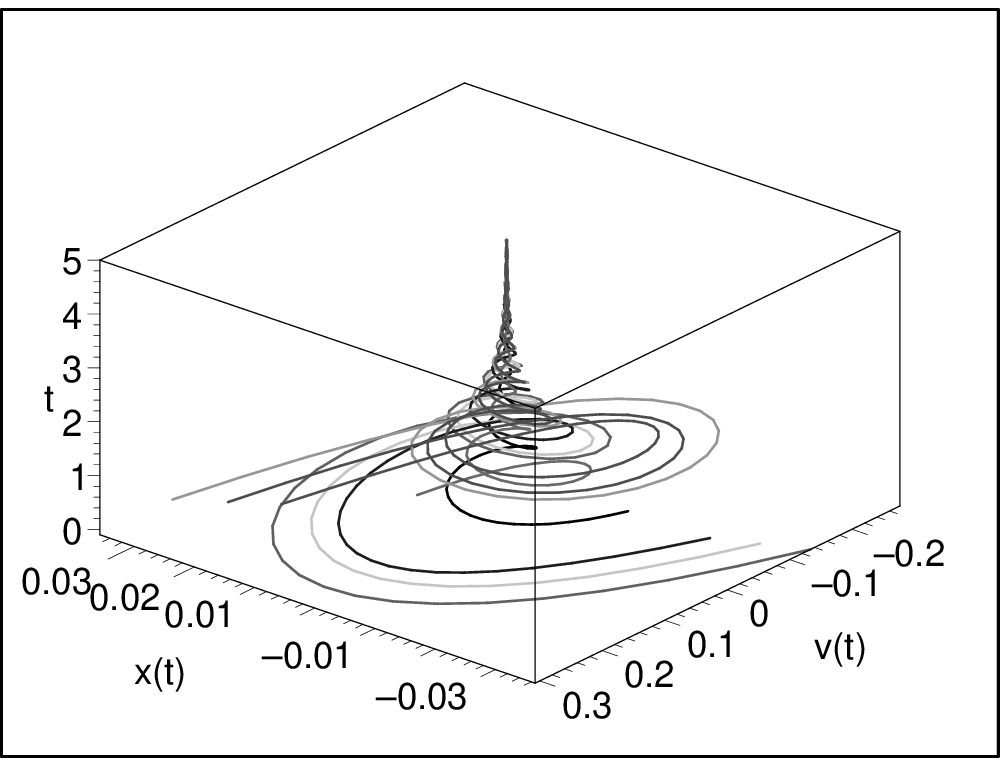}
\includegraphics[width=4cm,height=3.5cm]{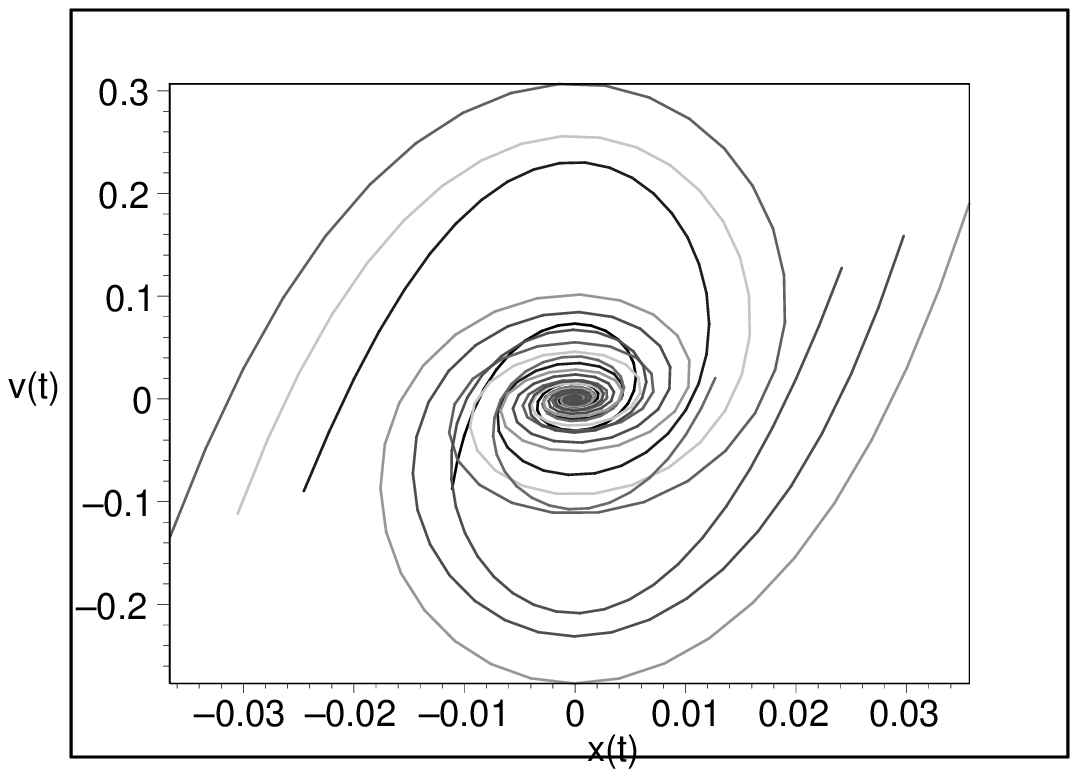}
\includegraphics[width=4cm,height=3.5cm]{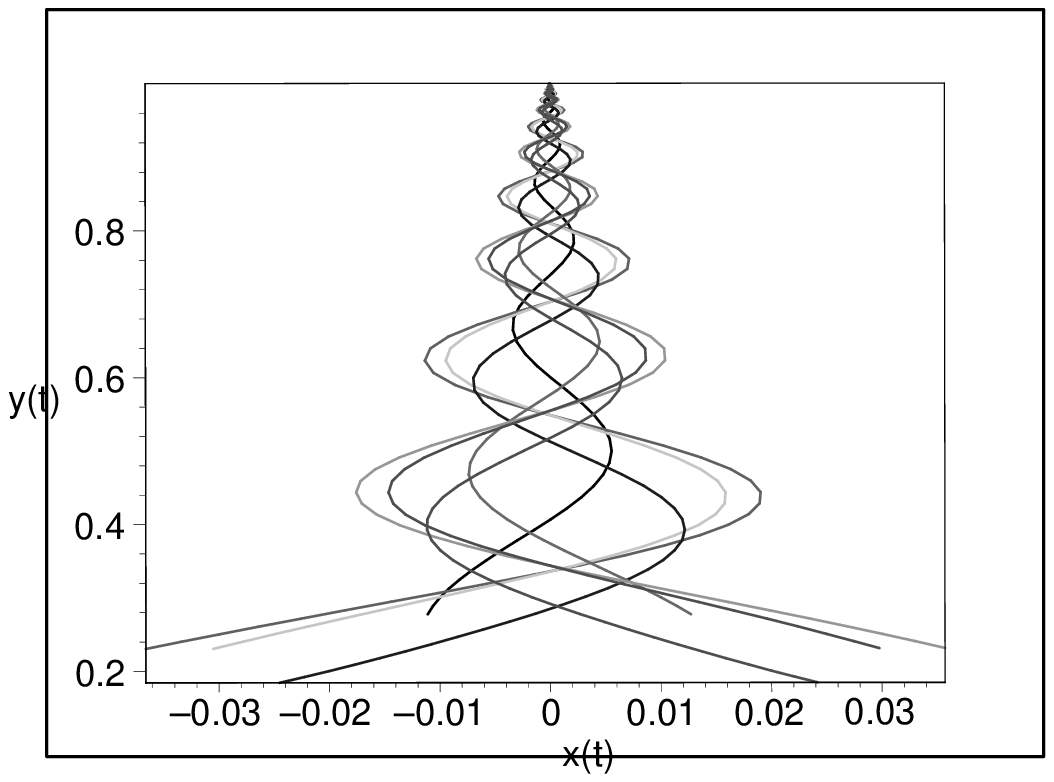}
\vspace{0.3cm}
\caption{Local behaviour of the orbits of the autonomous system of ODE
  (\ref{ode}) with function $f(v)$ in Eq.~(\ref{alphaleq1}), in the
  neightbourhood of the point $P_5$ associated with late-time cosmic
  dynamics, and the corresponding flux in time $\tau$ -- upper panels
  --, for the chosen values of the parameters: $\lambda=2$,
  $\alpha=0.1$ ($\gamma=4/3$-- background radiation), for different
  sets of initial conditions. In the lower panel we show the
  projection of the orbits into the phase planes $(x,v)$ and $(x,v)$,
  respectively.}\label{fig5}
\end{center}
\end{figure}

\begin{figure}[t!]
\begin{center}
\includegraphics[width=4cm,height=3.5cm]{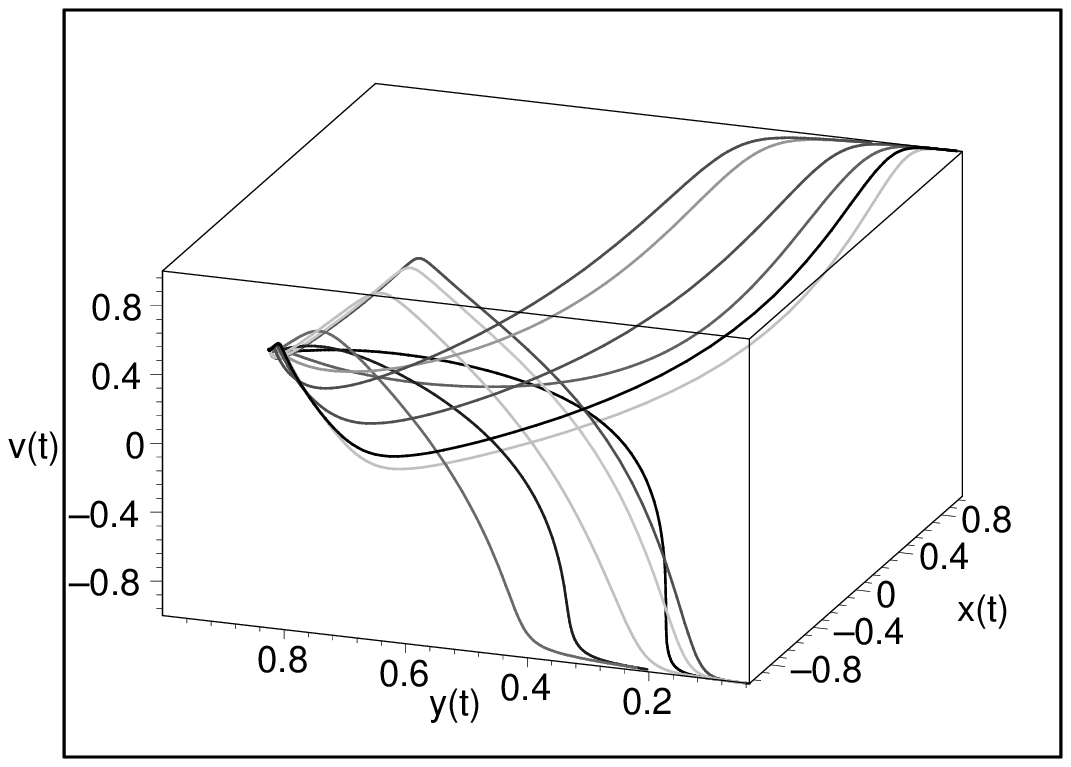}
\includegraphics[width=4cm,height=3.5cm]{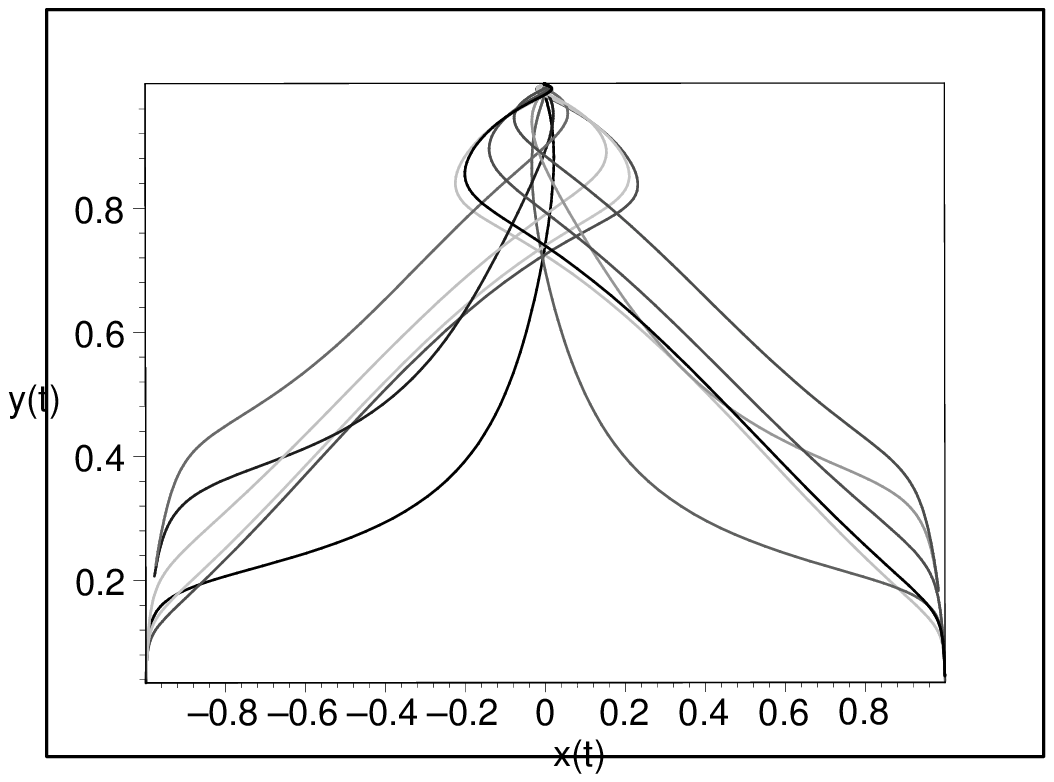}
\vspace{0.3cm}
\caption{Early times-to-intermediate behaviour of the orbits of the
  autonomous system of ODE~(\ref{ode}) with function $f(v)$ in
  Eq.~(\ref{alphaleq1}), for the chosen values of the parameters:
  $\lambda=2$, $\alpha=0.1$ ($\gamma=4/3$-- background radiation), for
  different sets of initial conditions. In the right-hand panel we
  show the projection of the orbits into the phase plane $(x,v)$. It
  is evident that the kinetic energy-dominated solution (point $P_2$
  in Tab.\ref{tab1}) is the past attractor, while the scaling solution
  (point $P_4=(0.81,0.71,1)$) is a saddle equilibrium
  point.}\label{fig6}
\end{center}
\end{figure}

For self-consistency, we show in Figs.~\ref{fig5} and~\ref{fig6} the
late-time (point $P_5$ above, associated with the neighbourhood of the
minimum of the potential), and early-to-intermediate times local
behaviours of the orbits of Eqs.~(\ref{ode}), respectively, for small
values of the parameter $\alpha$ ($0 < \alpha < 1$). 

Due to the choice of the values of the free parameters ($\lambda=2,
\alpha=0.1$), obeying $\lambda^2 > \alpha/4$ as in the former case,
since the corresponding equilibrium point is a spiral one, the orbits
coil around the de Sitter segment $(x,y,v)=(0,y,0)$ until, eventually,
they reach the de Sitter solution $3H = \Lambda$
(Fig.~\ref{fig5}). The early times-to-intermediate behaviour
(Fig.~\ref{fig6}), clearly shows that the kinet energy-dominated
solution (equilibrium point $P_2$) is the past attractor, while the
matter-scaling solution (point $P_4$) is a saddle equilibrium point.

We want to emphasize that the above spiral form of the orbits of the
ODE~(\ref{ode}) -- see Fig.~\ref{fig5} -- is what can be properly
correlated with CDM behavior, so that, only for $\lambda >
\sqrt{\alpha}/2$ the scalar field component in our model may be called
as SFDM.

\section{Summary of Results}
The relevant results of the study of the fixed points of the
autonomous system of ODE~(\ref{ode}), for the cosh-like potential
$V(\phi)$ in Eq.~(\ref{pot}) -- corresponding to $f(v)$ in
Eq.~(\ref{fv}) --, can be summarized as follows (see the former
section).

\begin{itemize}
\item The past (early-time) attractor in the phase space $\Psi$, for
  $\lambda < 6$, is the kinetic energy-dominated (decelerated) solution
  $3H^2 = \dot{\phi}^2/2$ (see Fig.\ref{fig3}). For values $\lambda \geq 6$
  there does not exist any past attractor in $\Psi$ (see
  Fig.\ref{fig4}).

\item The scalar field-dominated phase $3H^2 = \dot{\phi}^2/2 +
  V(\phi)$ (existing only for $\lambda^2 \leq 6$, and inflationary
  whenever $\lambda < \sqrt{2}$), and the scaling solution
  $\Omega_\phi/\Omega_\gamma =1/[(\lambda^2/ 3\gamma) -1]$ (exists
  whenever $\lambda^2 \geq 3\gamma$, decelerating if $\gamma\geq
  2/3$), are always saddle equilibrium points.

\item The (decelerating) matter-dominated solution $3H^2 =
  \rho_\gamma$ exists for any value $v$ in $\Psi$. It always
  represents a saddle fixed point (see Fig.~\ref{fig3}).

\item The future attractor is always the inflationary de Sitter
  solution $3H^2 = \Lambda$. At late times, for $\lambda >
  \sqrt{\alpha}/2$, the orbits of the ODE~(\ref{ode}) coil around the
  de Sitter segment $\{(0,y,0) : 0 \leq y \leq 1\}$ until, eventually,
  these reach the late-time de Sitter attractor $3H^2 = \Lambda$ (see
  Fig.~\ref{fig4}). The later (late-time) behaviour can be associated
  with oscillating solutions that reproduce CDM in the model.
\end{itemize}
In the next section we will discuss the relevance of the above results
to the cosmology of the model under investigation.

\section{Discussion and Conclusions}

The fact that the scaling solution (point $P_4$ in Table~\ref{tab1}),
the matter-dominated phase (point $P_1$), and the de Sitter
inflationary solution ($P_5$) are equilibrium points of
Eqs.~(\ref{ode}) for the potential~(\ref{pot}), is not a happy
coincidence but a profound principle of nature, inherent in the SFDM
model with cosh-like potential under investigation in this paper. 

In terms of the dynamical systems analysis this means, in particular,
that independent on the initial conditions chosen, there is a chance
for the cosmological model to transit by each of the mentioned phases,
spending some time in the neighbourhood of each one of
them.\footnote{The neccessary requirement for the occurrence of an
  equilibrium point $x^\prime = y^\prime = v^\prime =0$, means that
  the universe may keep evolving in the neighbourhood of this point
  for a quite long time, depending on the nature of the fixed point:
  if the equilibrium point is a saddle in the phase space, it is
  correlated with a transient stage of the cosmic evolution, but if it
  is an attractor, it may only represent either the starting point of
  the evolution -- past or early-time attractor --, or its end point
  -- future or late-time attractor.} More precisely, independent on
the initial conditions, the cosmological evolution might drive the
universe from one phase to another to, inevitably, end up in the de
Sitter stage (point $P_5$) being the late-time attractor in the phase
space of the model ($\Psi$).

The cosmic evolution driven by the $\phi$-field after the end of
inflation, most probably starts in the neightbourhood of $P_4$ where
the scalar field tracks matter (radiation). Then, at
early-to-intermediate times it approaches to the matter-dominated
(decelerated) stage $P_1$ where most structure was formed until,
finally, the evolution end ups in the inflationary de Sitter stage
$P_5$. 

The fact that, for appropriate values of the free parameters ($\lambda
> \sqrt{\alpha}/2$), the latter solution is a spiral equilibrium point
means that, at late times, while the phase space trajectories approach
to the de Sitter phase, the scalar field perturbations perform damped
oscillations that mimic CDM. Therefore the SFDM model with a cosh-like
potential could also be a nice scenario to address the united
description of the dark matter and of the dark energy in a single
field\cite{cul}.

\begin{acknowledgements}
This work was partly supported by CONACyT M\'exico (46195, 49865-F,
54576-F, 56159-F, 56946), PROMEP UGTO-CA-3 and CONACYT,
grant number I0101/131/07 C-234/07, Instituto Avanzado de Cosmologia
(IAC) collaboration. I.Q. aknowledges also the MES of Cuba for partial
support of the research. The numeric computations were carried out in
the "Laboratorio de Super-C\'omputo Astrof\'{\i}sico (LaSumA) del
Cinvestav". 
\end{acknowledgements}

\end{document}